\documentclass[11pt]{article}
\usepackage{moriond,epsfig}

\def\be{\begin{equation}}
\def\ee{\end{equation}}
\def\bea{\begin{eqnarray}}
\def\eea{\end{eqnarray}}

\begin{document}
\vspace*{4cm}
\title{Recent Results on $\psi(3770)$ Physics at BES}

\author{ M.G. ZHAO (For BES Collaboration) }

\address{Institute of High Energy Physics, Chinese Academy of Science,\\
Beijing 100049, P.R. China}

\maketitle\abstracts{ There are about 33, 6.5 and 1.0 pb$^{-1}$ of
$e^+e^-$ annihilation data have been taken around the
center-of-mass energies of $\sqrt s=$ 3.773 GeV, at $\sqrt s=$
3.650 GeV and at $\sqrt s=$ 3.6648 GeV, respectively, with the
BES-II detector at the BEPC collider.  By analyzing these data
sets, we have measured the branching fraction for $\psi(3770)\to$
non-$D\bar D$ by several different methods; and have observed an
anomalous line shape of $\sigma^{e^+e^-\to{\rm hadrons}}$ in
energy region from 3.65 to 3.87 GeV; and have measured the line
shapes of the $D^+D^-$, $D^0\bar D^0$ and $D\bar D$ production and
together with the ratios of the production rates of $D^+D^-$ and
$D^0\bar D^0$ in $e^+e^-$ annihilation around the $\psi(3770)$
resonance. }

\section{Introduction}
The $\psi(3770)$ is the lowest mass charmonium resonance above the
open charm pair $D\bar D$ production threshold. Traditional
charmonium theories expect that it decays almost entirely into
pure $D\bar D$ pairs. However, before the measurements from BES
and CLEO Collaborations, there is a long standing puzzle that the
observed cross section $\sigma^{\rm obs}_{\psi(3770)}$ for
$\psi(3770)$ production is not saturated by the observed cross
section $\sigma^{\rm obs}_{D\bar D}$ for $D\bar D$ production at
the $\psi(3770)$ peak \cite{0506051}. Recently, CLEO Collaboration
measured the $e^+e^-\to\psi(3770)\to{\rm non}-D\bar D$ cross
section to be $(-0.01\pm0.08^{+0.41}_{-0.30})$ nb
\cite{prl96_092002}. However, BES measured the branching fraction
for $\psi(3770)\to{rm non}-D\bar D$ by analyzing several different
data samples and different methods to be $(14.7\pm3.2)\%$
\cite{plb641_145,prl97_121801,prd76_122002,plb659_74,pdg2008} with
the assumption that there is only one traditional $\psi(3770)$
resonance in the energy region between 3.700 and 3.872 GeV. While,
up to now, the sum of the measured branching fractions for
$\psi(3770)\to$ exclusive non$-D\bar D$ decays is not more than
$2\%$
\cite{hepnp28_4_325,plb605_63,prl96_082004,prl96_182002,prd74_031106,prd74_012005}.
To better understand the situation, we examine the line shape of
the cross section for $e^+e^-\to{\rm hadrons}$ in the energy
region from 3.650 to 3.872 GeV, and measure the line shapes of the
$D^+D^-$, $D^0\bar D^0$ and $D\bar D$ production and the ratios of
the production rates of $D^+D^-$ and $D^0\bar D^0$ in $e^+e^-$
annihilation at $\psi(3770)$ resonance. These measurements are
made by analyzing about 33, 6.5 and 1.0 pb$^{-1}$ of $e^+e^-$
annihilation data sets taken around the center-of-mass energies of
$\sqrt{s}$ = 3.773 GeV, at $\sqrt{s}$ = 3.650 GeV and at
$\sqrt{s}$ = 3.6648 GeV with the BES-II detector at the BEPC
collider. The data sets taken around $\sqrt{s}$ = 3.773 GeV
include the data sets of 17.3 pb$^{-1}$ taken at $\sqrt{s}$ =
3.773 GeV, the precision cross section scan data sets taken in
March 2001, during March to April 2003 and during December 2003 to
January 2004.

\section{Measurements about $\mathcal{B}(\psi(3770)\to {\rm non}-D\bar D)$}
Assuming that there is only one traditional $\psi(3770)$ in the
energy region between 3.700 and 3.872 GeV, we measure the
branching fraction for $\psi(3770)\to{rm non}-D\bar D$ decays by
analyzing several different data samples and different methods.
The measured branching fractions for $\psi(3770)\to$ $D^0\bar
D^0$, $D^+D^-$, $D\bar D$ and non$-D\bar D$ decays are compared in
Tab. \ref{tab::nondd}. After the $\psi(3770)$ resonance was
discovered for more than thirty years, Particle Data Group 2008
\cite{pdg2008} gives the branching fractions for $\psi(3770)\to$
$D^0\bar D^0$, $D^+D^-$ and $D\bar D$ decays for the first time.
They are $\mathcal{B}[\psi(3770)\to D^0\bar D^0] =
(48.7\pm3.2)\%$, $\mathcal{B}[\psi(3770)\to D^+D^-] =
(36.1\pm2.8)\%$ and $\mathcal{B}[\psi(3770)\to D\bar D] =
(85.3\pm3.2)\%$, which indicates that the branching fraction for
$\psi(3770)\to{\rm non}-D\bar D$ decays is $(14.7 \pm 3.2)\%$.

These measurements imply that the $\psi(3770)$ could substantially
decay into non$-D\bar D$ final states, which might greatly
challenge the traditional theories. Otherwise, there may exist
some other effects in the energy region around $\psi(3770)$
resonance which are responsible for the large branching fraction
for $\psi(3770)\to{\rm non}-D\bar D$ decays.

\begin{table}
\caption{The measured branching fractions for $\psi(3770)\to$
$D^0\bar D^0$, $D^+D^-$, $D\bar D$ and non$-D\bar D$ decays.
\label{tab::nondd}} \vspace{0.4cm}
\begin{center}
\begin{tabular}{|c|c|c|c|c|}
\hline $\mathcal{B}[\psi(3770)\to]$(\%)& $D^0\bar D^0$ &$D^+D^-$ &
$D\bar D$ & ${\rm non}-D\bar D$ \\ \hline Ref.\cite{plb641_145}&
$49.9 \pm 1.3 \pm 3.8$& $35.7 \pm 1.1 \pm 3.4$& $85.5 \pm 1.7 \pm
5.8$ & $14.5 \pm 1.7 \pm 5.8$ \\  Ref.\cite{prl97_121801}& $46.7
\pm 4.7 \pm 2.3$& $36.9 \pm 3.7 \pm 2.8$& $83.6 \pm 7.3 \pm 4.2$ &
$16.4 \pm 7.3 \pm 4.2$ \\ Ref.\cite{prd76_122002}& &
& & $13.4 \pm 5.0 \pm 3.6$ \\
Ref.\cite{plb659_74}& & & & $15.1 \pm 5.6 \pm 1.8$ \\\hline
PDG 2008\cite{pdg2008}&$48.7\pm3.2$ & $36.1\pm2.8$&$85.3\pm3.2$ & $14.7 \pm 3.2$ \\
\hline
\end{tabular}
\end{center}
\end{table}

\section{Anomalous line-shape of $\sigma^{e^+e^-\to{\rm hadrons}}$
in energy region from 3.65 to 3.87 GeV} To understand why the
measured branching fraction for  $\psi(3770)\to{\rm non}-D\bar D$
is substantially larger than 2\%, we examine the line shape of the
cross section for $e^+e^-\to$ hadrons in the energy region from
3.650 to 3.872 GeV, by analyzing the precision cross section scan
data sets taken in March 2003 and during December 2003 to January
2004. Fig. 2 in Ref. \cite{prl101_102004} shows the measured
observed cross sections for $e^+e^-\to$ hadrons versus the nominal
center-of-mass energies. In the figure, we can see that the slope
of the high-energy side of the peak is substantially larger than
that of the low-energy side. This phenomenon is inconsistent with
the traditional expectation under the assumption that there is
only one simple $\psi(3770)$ resonance in this energy region. To
investigate this situation, we fit the measured observed cross
sections for $e^+e^-\to $ hadrons with the following solutions,
respectively. Firstly, we suppose that there are two amplitudes
and ignore the possible interference between them. Secondly, we
suppose that there are two amplitudes completely interfering with
each other. Thirdly, we assume that there are two amplitudes of
G(3900) \cite{prd76_111105r,prd77_011103r} and $\psi(3770)$
resonance interfering with each other. Finally, we carry out the
treatment as a comparison solution that there is only one simple
$\psi(3770)$ resonance. The fitted results are summarized in Tab.
\ref{tab::anomalous}. The details about the fits can be found in
Ref. \cite{prl101_102004}. By comparing the fitted results, we can
obtain the better hypothesis to describe the anomalous line shape
of the cross sections for $e^+e^-\to$ hadrons in the energy region
from 3.700 to 3.872 GeV. The signal significance for the two
structure hypotheses are $7.0\sigma$ and $7.6\sigma$ for solution
1 and solution 2. The significance of the interference between the
two amplitudes is $3.6\sigma$. Fig. 2 in Ref. \cite{prl101_102004}
shows the fit to the observed cross sections for $e^+e^-\to$
hadrons for solution 2. Fig. 3 (a) in Ref.\cite{prl101_102004}
shows the fits to the observed cross sections for $e^+e^-\to$
hadrons for the three solutions. Fig. 3 (b) in
Ref.\cite{prl101_102004} shows the ratio of the residual between
the observed cross section and the fitted value for the one
$\psi(3770)$ amplitude hypothesis to the error of the observed
cross section, which indicating that there is more likely some new
structure in addition to $\psi(3770)$ resonance.

\begin{table}
\caption{The fitted results, where $M$, $\Gamma^{\rm tot}$ and
$\Gamma^{\rm ee}$ are the mass, total, and leptonic widths of
resonance(s), $\sigma_G$ is standard deviation of $G(3900)$,
$\phi$ is the phase difference between the two amplitudes and {\bf
AM} stands for amplitude(s). ndof denotes number of degrees of
freedom. \label{tab::anomalous}} \vspace{0.4cm}
\small
\begin{center}
\begin{tabular}{|c|c|c|c|c|}
\hline Quality & two {\bf AM} (solution 1) & two {\bf AM} (solution 2)
& one {\bf AM} & $\psi(3770)$ and $G(3900)$  \\
  &  &  &  & {\bf AM} (solution 3)
\\
\hline $\chi^2$/(ndof) & 125/103=1.21 & 112/102=1.10 &
182/106=1.72 & 170/104=1.63 \\
$M_{\psi(3686)}$ [MeV]& $3685.5\pm0.0\pm0.5$ & $3685.5\pm0.0\pm0.5$
& $3685.5\pm0.0\pm0.5$ & $3685.5\pm0.0\pm0.5$ \\
$\Gamma^{\rm tot}_{\psi(3686)}$ [keV]& $312\pm34\pm1$
& $311\pm38\pm1$ & $304\pm36\pm1$ & $293\pm36\pm1$ \\
$\Gamma^{\rm ee}_{\psi(3686)}$ [keV]& $2.24\pm0.04\pm0.11$
& $2.23\pm0.04\pm0.11$ & $2.24\pm0.04\pm0.11$ & $2.23\pm0.04\pm0.11$ \\
$M_{1}$ [MeV]& $3765.0\pm2.4\pm0.5$ & $3762.6\pm11.8\pm0.5$
& $3773.3\pm0.5\pm0.5$ & $3774.4\pm0.5\pm0.5$ \\
$\Gamma^{\rm tot}_{1}$ [MeV]& $28.5\pm4.6\pm0.1$
& $49.9\pm32.1\pm0.1$ & $28.2\pm2.1\pm0.1$ & $28.6\pm2.3\pm0.1$ \\
$\Gamma^{\rm ee}_{1}$ [eV]& $155\pm34\pm8$
& $186\pm201\pm8$ & $260\pm21\pm8$ & $264\pm23\pm8$ \\
$M_{2}$ [MeV]& $3777.0\pm0.6\pm0.5$ & $3781.0\pm1.3\pm0.5$
& ... & 3943.0(fixed) \\
$\Gamma^{\rm tot}_{2}$ [MeV]& $12.3\pm2.4\pm0.1$
& $19.3\pm3.1\pm0.1$ & ... & ... \\
or $\sigma_G$ [MeV]& ...
& ... & ... & 54(fixed) \\
$\Gamma^{\rm ee}_{2}$ [eV]& $93\pm26\pm9$
& $243\pm160\pm9$ & ... & ... \\
or C & ... & ... & ... & 0.243(fixed) \\
$\phi$ [degree] & ... & $158\pm334\pm5$ & ... & $150\pm23\pm5$ \\
f & $0.4\pm5.6\pm0.6$ & $5.2\pm2.5\pm0.6$ & $0.0\pm0.5\pm0.6$ & $0.0\pm1.2\pm0.6$ \\
\hline
\end{tabular}
\end{center}
\end{table}




\section{Measurement of the line shapes of $sigma^{D\bar D}(s)$
and $sigma^{D^+D^-}(s)/sigma^{D^0\bar D^0}$ around $\psi(3770)$
resonance} To investigate what on earth is responsible for the
large branching fraction for $\psi(3770)\to{\rm non}-D\bar D$
decays which is beyond the expectation by the traditional
theories, we measure the line shapes of the $D^0\bar D^0$,
$D^+D^-$, and $D\bar D$ production and the ratios of the
production rates of $D^0\bar D^0$ and $D^+D^-$ in $e^+e^-$
annihilation at $\psi(3770)$ resonance. These measurements are
also helpful for the understanding the anomalous line shape of the
cross section for $e^+e^-\to$ hadrons in the energy region from
3.650 to 3.872 GeV. These measurements are made by analyzing the
precision cross section data sets taken in March 2001, during the
period from March to April 2003, and during December 2003 to
January 2004. Fig. 5 in Ref.\cite{plb668_263} shows the observed
cross sections for $e^+e^-\to D^0\bar D^0$, $e^+e^-\to D^+D^-$ and
$e^+e^-\to D\bar D$ versus the nominal center-of-mass energies. In
the figure, we can see that the line shapes of the cross section
for $e^+e^-\to D^0\bar D^0$, $e^+e^-\to D^+D^-$ and $e^+e^-\to
D\bar D$ are also anomalous, just like the line shape of the
observed cross section for $e^+e^-\to$ hadrons. Fig. 6 in
Ref.\cite{plb668_263} shows the measured ratio of the observed
cross section for $e^+e^-\to D^+D^-$ relative to the observed
cross section for $e^+e^-\to D^0\bar D^0$ versus the nominal
center-of-mass energies.



\section{Summary}
Using the $e^+e^-$ data sets of about 33, 6.5 and 1.0 pb$^{-1}$,
respectively, taken around the center-of-mass energies of
$\sqrt{s}$ = 3.773 GeV, at $\sqrt{s}$ = 3.650 GeV and at
$\sqrt{s}$ = 3.6648 GeV with the BES-II detector at the BEPC
collider, BES Collaboration measure the branching fraction for
$\psi(3770)\to{\rm non}-D\bar D$ decays by analyzing several
different data samples and different methods with assumption that
there is only one traditional $\psi(3770)$ in the energy region
between 3.700 and 3.872 GeV. We observe an anomalous line shape of
the cross section for $e^+e^-\to$ hadrons in the energy region
from 3.650 to 3.872 GeV. We measure the line shapes of the
$D^+D^-$, $D^0\bar D^0$ and $D\bar D$ production and the ratio of
the production rates of $D^+D^-$ and $D^0\bar D^0$ in $e^+e^-$
annihilation at $\psi(3770)$ resonance. These indicate that there
may exist a new structure in addition to one simple $\psi(3770)$
resonance in the energy region between 3.700 and 3.872 GeV or
there are some unknown dynamics effects distorting the line shape
of the cross sections for $e^+e^-\to$ hadrons and $D\bar D$.

\section*{Acknowledgments}
The BES collaboration thanks the staff of BEPC for their hard
efforts. This work is supported in part by the National Natural
Science Foundation of China under contracts Nos. 10491300,
10225524, 10225525, 10425523, the Chinese Academy of Sciences
under contract No. KJ 95T-03, the 100 Talents Program of CAS under
Contract Nos. U-11, U-24, U-25, the Knowledge Innovation Project
of CAS under Contract Nos. U-602, U-34 (IHEP), the National
Natural Science Foundation of China under Contract No. 10225522
(Tsinghua University).

\end{document}